\documentclass[prb,aps,showpacs,superscriptaddress,amsmath,twocolumn,floatfix]{revtex4}
\usepackage{graphicx}
\usepackage{multirow}
\usepackage{color}
\newcommand{\br}{\mathbf{r}} \newcommand{\brp}{\mathbf{r'}}
\newcommand{\bk}{\mathbf{k}} \newcommand{\bkp}{\mathbf{k'}}
\newcommand{\bq}{\mathbf{q}} 
\newcommand{\bG}{\mathbf{G}} 
\newcommand{\bR}{\mathbf{R}} 
\newcommand{\bz}{\mathbf{0}} 

\begin{document}
\title{Efficient and accurate calculation of exact exchange and RPA
correlation energies in the Adiabatic-Connection Fluctuation-Dissipation
theory}
\author{Huy-Viet Nguyen}
\affiliation{SISSA-Scuola Internazionale Superiore di Studi Avanzati, via Beirut 2-4, I-34014 Trieste, Italy}
\affiliation{Physics Faculty, Hanoi National University of Education, 136 Xuan-Thuy, Cau-Giay, Hanoi, Vietnam}
\author{Stefano de Gironcoli}
\affiliation{SISSA-Scuola Internazionale Superiore di Studi Avanzati, via Beirut 2-4, I-34014 Trieste, Italy}
\affiliation{CNR-INFM DEMOCRITOS National Simulation Centre, via Beirut 2-4, I-34014 Trieste, Italy}
\date{\today}
\begin{abstract}
Recently there has been a renewed interest in the calculation of
exact-exchange and RPA correlation energies for realistic systems.
These quantities are main ingredients of the so-called EXX/RPA+
scheme which has been shown to be a promising alternative approach to
the standard LDA/GGA DFT for weakly bound systems where 
LDA and GGA perform poorly. In this paper, we present an efficient
approach to compute the RPA correlation energy in the framework of the
Adiabatic-Connection Fluctuation-Dissipation formalism. The method is
based on the calculation of a relatively small number of eigenmodes of
RPA dielectric matrix, efficiently computed by iterative density response
calculations in the framework of Density Functional Perturbation Theory. We
will also discuss a careful treatment of the integrable divergence in
the exact-exchange energy calculation which alleviates the problem of
its slow convergence with respect to Brillouin zone sampling. As an
illustration of the method, we show the results of applications to bulk
Si, Be dimer and atomic systems.

\end{abstract}
%
%
\maketitle
\section{Introduction}
Developments during the last several decades have brought Density
Functional Theory (DFT) to be the method-of-choice in many calculations of
physical and chemical properties from first-principles. Despite its many
spectacular successes, Kohn-Sham (KS) density-functional theory within
the LDA and GGAs approximations for the exchange-correlation (xc) functional still
has some well-known drawbacks. This includes the failure of standard DFT
functionals in the description of van der Waals systems where long-range
correlation effects are predominant. This failure manifests clearly
in erratic or completely failed results when LDA/GGA DFT is applied
to sparse matter, including layered structures, polymer and molecular
crystals, or weakly-bound chemical systems.

Of all the attempts to overcome these drawbacks, the approach based on
the formally exact expression of xc energy in
term of linear response functions derived from the Adiabatic Connection
Fluctuation-Dissipation (ACFD) theorem\cite{ACFDT} provides a promising
way to develop a systematic improvement.

The method is in principle entirely parameter free--although an
appropriate definition of the xc kernel is still needed--and a few
difficult cases where standard DFT fails qualitatively have been described
with satisfactory results.\cite{Fuchs,Marini,Harl} Practical applications in a
few case-studies have shown that not only the description of van der
Waals systems is improved but that the already satisfactory description
of many molecules and solids are also not compromised.\cite{Marini,Garcia}
The method is very computationally demanding and this has prevented
its direct application to complex systems. Most often it is
limited to a post-self-consistent correction where the accurate
exchange-correlation energy is computed from the charge density obtained
from a self-consistent calculation performed with a more traditional
xc functional. The ACFD formalism also serves as the starting point
for further simplifications that have led to the development of an
approximate vdW-DF functional by Langreth and coworkers,\cite{vdW-DF}
which allows for the treatment of large systems and has recently been
made fully self-consistent.\cite{vdW-DF-scf}

We will address here the computational difficulties involved in the full
evaluation of ACFD exchange-correlation energies of realistic systems,
presenting an efficient general formulation of the problem and its
practical implementation in the plane-wave pseudopotential (PW PP)
formalism.

Previous implementations of ACFD formulas rely on the construction of
the full Kohn-Sham response function from the spectrum of the KS
Hamiltonian. Basic definitions and some details are given in 
Sec.~\ref{Theory}.
A more efficient scheme is then presented in Sec.~\ref{ACFD} that avoids the full
diagonalization of the KS Hamiltonian and the cumbersome summation over
occupied and unoccupied states needed in the traditional calculation
of the response functions, as well as the computationally demanding
operations involved in the solution of the Dyson equation via exact
algebra. These steps are replaced by the iterative calculation of a relatively
small number of eigenmodes of the RPA dielectric matrix as explained in
its general formulation in the first part of this section. 
Some technical details of the implementation in the PW PP formalism and
a careful treatment of the integrable divergence which alleviates the
problem of slow convergence with respect to Brillouin zone sampling
in the calculation of the exact-exchange energy are also discussed 
here.
To demonstrate the efficiency of the implementation, we will present in
Sec.~\ref{App} results of applications to selected paradigmatic
systems, namely bulk Si, weakly bound Be dimer, and a number of spherical
atomic systems.
\section{Theory} \label{Theory}
We recall here some basic definitions in ACFD theory as well as some
computational details of currently available PW PP implementations.
\subsection{Exchange and correlation energy in ACFD theory}
In the ACFD formalism, exchange-correlation energy, $E_{xc}$, of an electronic 
system with density $n(\mathbf{r})$ can be written in the formally exact form
\begin{eqnarray}
E_{xc}=&-&\frac{\hbar}{2\pi}\int_0^1 d\lambda \int d\mathbf{r} d\mathbf{r'}
\frac{e^2}{|\mathbf{r}-\mathbf{r'}|}\nonumber\\
&\times&\left\{\int_0^{\infty} du\chi_{\lambda}(\mathbf{r},\mathbf{r'};iu)+
\delta(\mathbf{r}-\mathbf{r'})n(\mathbf{r}) \right\},
\end{eqnarray}
where $\chi_{\lambda}(\mathbf{r},\mathbf{r'};iu)$ is the density response
function at imaginary frequency, $iu$, of the system where Coulomb
interaction between electrons is scaled by $\lambda$, i.e. $\lambda
e^2/|\mathbf{r}-\mathbf{r'}|$, and the external potential is modified
so that the electronic density is the same as in the ground state of
the physical system ($\lambda=1$). The exchange-correlation energy can
be furthermore separated into the KS exchange energy, $E_x$, and the
correlation energy, $E_c$. The former is the counterpart of Hartree-Fock
exchange energy in the context of density-functional theory. It is
given by the well-known expression in term of occupied KS orbitals
(a non magnetic system is considered for simplicity and a factor of two
accounts for spin degeneracy)
\begin{equation}
E_x=- 2 \frac{e^2}{2}\int d\mathbf{r}d\mathbf{r'}\frac{| \sum_{i}^{occ}\phi_i^{\ast}(\mathbf{r})
\phi_i(\mathbf{r'})|^2}{|\mathbf{r}-\mathbf{r'}|}.
\label{EXX}
\end{equation}
The latter can be expressed in a compact form in terms of 
linear density responses, in matrix notation, as
\begin{equation}
E_c = -\frac{\hbar}{2\pi}\int_0^1 d\lambda\int_0^{\infty} du~ 
Tr\left\{v_c[\chi_{\lambda}(iu)-\chi_0(iu)]\right\}.
\label{Ec}
\end{equation}
In Eq. (\ref{Ec}), $v_c$ is the electron-electron interaction kernel,
$e^2/|\mathbf{r}-\mathbf{r'}|$, and $\chi_0(iu)$ is the density
response function of the non-interacting electron system and can be
explicitly expressed in term of, occupied and empty, KS orbitals,
$\phi_i(\mathbf{r})$, KS eigenvalues, $\epsilon_i$, and occupation
numbers, $f_i$,
\begin{equation}
\chi_0(\mathbf{r},\mathbf{r'};iu)=2\sum_{i,j}(f_i-f_j)
\frac{\phi_i^{\ast}(\mathbf{r})\phi_j(\mathbf{r})\phi_j^{\ast}(\mathbf{r'})
\phi_i(\mathbf{r'})}{\epsilon_i-\epsilon_j+i\hbar u},
\label{chi0}
\end{equation}
where again a factor of two is present accounting for spin degeneracy.
For $\lambda > 0$, the interacting response function, $\chi_{\lambda}$, is
related to $\chi_0$ by a Dyson-like equation with an exchange-correlation
kernel, $f_{\lambda}^{xc}(iu)$, which is still unknown:
\begin{equation}
\chi_{\lambda} = \chi_{0} + \chi_{0}[\lambda v_c+f_{\lambda}^{xc}]\chi_{\lambda}.
\label{Dyson}
\end{equation}
In the above formula, frequency dependence is implied in each term (with
the exception of the Coulomb kernel, $v_c$) and spatial coordinate
dependence is implicit in the matrix notation.
Despite the exactness of the expressions presented above, one needs to
use an approximation for the xc-kernel in practical applications. When
the xc-kernel is specified, the system of Eqs. (\ref{Ec})$-$(\ref{Dyson})
is closed and allows one to evaluate the  correlation energy.
\subsection{Total energy in EXX/RPA+ scheme }
Experience has shown that none of the available approximate xc-kernels 
gives a systematic improvement; different kernels perform better for different systems.
Although the RPA-kernel, i.e. simply setting $f_{\lambda}^{xc}$ in
Eq.~(\ref{Dyson}) to zero, is exact for long-range correlation, it is a
poor approximation for short-range correlation. When applied to atomic
or molecular systems, RPA correlation energy alone is very inaccurate.
In fact, it was abandoned decades ago as a DFT approximate xc functional
until a quite recent work by Kurth and Perdew\cite{Kurth} showed 
that it is possible to correct the deficiency of RPA in a simple way by
combining the full RPA with a local- or semilocal-density correction for
short-range correlation. This leads to the so-called RPA+ correlation energy 
in which the local-density correction for the short-range correlation is 
chosen in such a way that $E_{\textrm{c}}^{\textrm{RPA+}}$ becomes exact in the
limit of homogeneous electron gas (HEG)
\begin{equation}
E_{\textrm{c}}^{\textrm{RPA+}}=E_{\textrm{c}}^{\textrm{RPA}} - 
(E_{\textrm{c}}^{\textrm{LDA-RPA}}-E_{\textrm{c}}^{\textrm{LDA}}).
\end{equation}
In the above definition, $E_{\textrm{c}}^{\textrm{LDA-RPA}}$ is
the local-density approximation of RPA correlation energy which is
exactly canceled out by $E_{\textrm{c}}^{\textrm{RPA}}$ in the limit of an
HEG. 

In order to have exchange and correlation energies treated on the same
footing, exchange energy must also be computed in the same formalism,
i.e. from Eq.~(\ref{EXX}).

In practice this EXX/RPA+ scheme is applied as a post-scf correction
and the energy is calculated by first performing a standard LDA/GGA
DFT calculation and then replacing the xc-energy at LDA level by
exact-exchange and RPA+ correlation energies calculated from the LDA/GGA
charge density.  In spite of this, the EXX/RPA+ scheme has been shown
to be a promising approach for the description of weakly bound systems
where standard LDA/GGA performs poorly.\cite{Fuchs,Marini,Harl}

\subsection{Existing implementations in PW PP method}
Existing implementations~\cite{Furche} of the ACFD formulas in plane-wave
pseudopotential computer codes start by first diagonalizing the KS
Hamiltonian for all the occupied and unoccupied KS orbitals,~\cite{Harl}
or at least a good part of the unoccupied states,\cite{Fuchs,Marini} so
that the KS response function $\chi_0(iu)$ can be calculated explicitly
according to its definition in Eq. (\ref{chi0}). The calculation then
proceeds by solving the Dyson-like equation for $\chi_{\lambda}(iu)$ for
a number of values of the coupling constant, $\lambda$, and of the imaginary frequency,
$iu$, and then by integrating over these variables to ultimately obtain
the correlation energy. An obvious disadvantage of these implementations
is that many unoccupied states need to be considered in order to get
well-converged results. This forces one to solve the KS problem using
full matrix diagonalization algorithms with unfavorable scaling instead
of very efficient iterative-diagonalization techniques, commonly used
to calculate the occupied KS orbitals in typical self-consistent
calculations. Moreover the summation over valence and conduction bands and
over all the k-points in the Brillouin zone for setting up the response
function (Eq. (\ref{chi0})) has been shown to be a very cumbersome
operation which prevents the application of the method to large systems.
\section{Efficient calculation of ACFD formulas}
\label{ACFD}

We describe now our method, first focusing on its general aspects, valid
for any symmetry and basis set, and then specializing the discussion to
a PW PP approach.

\subsection{General aspects and PW PP implementation}
\label{general}
Let us starts by defining the following generalized eigenvalue problems 
%
\begin{eqnarray}
\chi_0|w_i\rangle &=& a_iv_c^{-1}|w_i\rangle, \label{eigenchi0}\\
\chi_{\lambda}|z^{\lambda}_i\rangle &=& 
b^{\lambda}_iv_c^{-1}|z^{\lambda}_i\rangle, 
\label{eigenchi}
\end{eqnarray}
where $\chi_0, \chi_{\lambda}$ and $v_c$ are linear response 
and Coulomb operators in matrix notation, $\{|w_i\rangle, a_i\}$
and $\{|z^{\lambda}_i\rangle, b^{\lambda}_i\}$ are eigenpairs (all
these quantities depend implicitly on the imaginary frequency $iu$).
These eigenvalues problems are well-defined since all the operators are
symmetric and positive ($v_c$) or negative ($\chi_0, \chi_{\lambda}$)
definite. Once solutions of the generalized eigenvalue problems are
available, the traces of $v_c\chi_{\lambda}$ and $v_c\chi_{0}$ in
Eq. (\ref{Ec}) are simply evaluated by summing up all their (relevant)
eigenvalues. Working with these generalized eigenvalue problems has
several advantages when RPA approximation is considered:
(i) the set of formally $\lambda$-dependent eigenpotentials,
$\{|z^{\lambda}_i\rangle\}$, is actually $\lambda$-independent and
coincides with its non-interacting counterpart,  $\{|w_i\rangle\}$,
(ii) the corresponding eigenvalues are trivially related as 
\begin{equation}
b^{\lambda}_i = \frac{a_i}{1-\lambda a_i},
\end{equation}
which (iii) allows one to perform the $\lambda$ coupling constant integration in
Eq. (\ref{Ec})  analytically leading to the final expression
\begin{equation}
E_c = \frac{\hbar}{2\pi}\int_0^{\infty} du~\sum_i\{a_i(iu)+\ln(1-a_i(iu))\}.
\label{Ec2}
\end{equation}

The spectrum of the eigenvalue problem in Eq.~(\ref{eigenchi0}) is
bounded from above by zero, which is also an eigenvalue, corresponding
to a constant eigenpotential.  Note that the eigenvalues $\{a_i\}$
are closely related to those of the RPA dielectric matrix since
$\epsilon_{RPA}=1-v_c\chi_0$, and the calculation of $\{a_i\}$ is very
similar to the calculation of dielectric band structures introduced
a few decades ago.\cite{dielecband,Car81}  In principle, the full
spectra must be known in order to calculate the correlation energy
from Eq.~(\ref{Ec2}) above. In practice, however, previous calculations
\cite{dielecband,Car81,Hybertsen} have shown that only a small number of
eigenvalues of the RPA dielectric matrix significantly differs from unit.
This translates in the fact that only a small fraction of the eigenvalues
$\{a_i\}$ will be significantly different from zero and needs to be
explicitly included in the correlation energy sum; all the rest being so close
to zero that their contributions to $E_c$ can be treated in a suitable
approximation or even simply neglected.

Evaluation of low-lying eigenvalues of the non-interacting problem in
Eq. (\ref{eigenchi0}) is done efficiently by iterative diagonalization
procedure starting from a set of trial eigenpotentials.  The basic
operation involved in the iterative diagonalization is the calculation
of the non-interacting response to a trial potential, $\Delta n = \chi_0
\Delta v$. This is done resorting to the linear response techniques of
Density Functional Perturbation Theory,\cite{DFPT} simply generalized
to imaginary frequency.  

Technically, for any given non-interacting perturtbing potential, 
$\Delta v$, the induced charge density variation is 
\begin{equation}
\Delta n(\br) = 2 {\tt Re} \sum_i^{occ} \phi^*_i(\br) \Delta \phi_i(\br),
\end{equation}
where the sum runs over the occupied (valence) states and $\Delta
\phi_i(\br)$ is the (conduction-band projected) variation of the
single-particle state, $\phi_i(\br)$, that can be obtained as the solution
of the linear system:
\begin{equation}
\left[ H_{KS} +\alpha P_v - (\epsilon_i + i u) \right] |\Delta \phi_i \rangle =
- (1 - P_v) \Delta v |\phi_i\rangle,
\end{equation}
where $P_v = \sum_i^{occ} |\phi_i\rangle \langle \phi_i| $ is the
projector on the occupied manifold and $\alpha$ is a positive constant,
larger than the valence bandwidth, ensuring that the linear system is not
singular even in the limit for $i u \rightarrow 0$. 
The non-hermitian
linear systems are solved by employing a fast and smoothly converging
variant of the bi-conjugate gradient algorithm.\cite{Bi-CGSTAB}

Overall the procedure amounts to solving the usual
DFPT linear systems\cite{DFPT} in which the ground state valence eigenvalues
are shifted by the complex constant $iu$, as already explained by
Mahan\cite{Mahan80} long time ago for the case of atomic polarizabilities.
Note however that, at variance with standard DFPT, no self-consistent
cycle is needed to obtain $\Delta n$ as the response of the
non-interacting system is being considered here.

As in static DFPT calculations, only occupied Kohn-Sham orbitals and
their linear response need to be calculated, instead of the full KS
spectrum needed in other implementation.\cite{Fuchs,Marini,Harl}

The scheme describe above have been implemented in the {\em Quantum
ESPRESSO} distribution.\cite{espresso}
Very similar ideas have also been recently reported by Wilson and
coworkers\cite{Wilson} for the calculation of dominant eigenpotentials
of static dielectric matrices.
%
%

The description presented so far is to a large extent basis set independent.
For periodic systems, matrix representations of the Kohn-Sham linear response
functions satisfy the Bloch theorem and can be classified by a vector
$\mathbf{q}$ in the Brillouin zone. The generalized eigenvalue problem
in Eq.~(\ref{eigenchi0}) can therefore be solved separately for each
$\mathbf{q}$ vector.  As a consequence, the expression for the correlation
energy per unit cell becomes
\begin{equation}
E_c = \frac{\hbar}{2\pi}\int_0^{\infty}du~\frac{1}{N_q}\sum_{q=1}^{N_q}
\sum_{i}^{N_{eig}}\{a_i(\mathbf{q}, iu)+\ln(1-a_i(\mathbf{q}, iu))\},
\label{Ec2k}
\end{equation}
where the notation $\frac{1}{N_q}\sum_{q=1}^{N_q}$ indicates the average
over the Brillouin zone and can be performed using the special point
technique that takes advantage of the point group symmetry of the system.

Special care is required in the limit of $\mathbf{q} \rightarrow 0$
since the leading matrix elements $\chi_0^{00}(\mathbf{q} \rightarrow 0)$
and $v_c^{-1}$ both go to zero as $|\mathbf{q}|^2$ making the eigenvalue
problem (\ref{eigenchi0}) ill-defined.
A simple solution to the problem is to use shifted grids of $\mathbf{q}$ 
vectors in the special-point summation.

\subsection{Exchange energy}
\label{exx_extra}
Although the main focus of this work is the calculation of correlation
energy it is clear that an accurate evaluation of the exchange term is
also necessary. A subtle problem, not widely recognized in the literature,
related to the $\mathbf{q} \rightarrow 0$ divergence of the Coulomb
interaction is present in the calculation of exact-exchange energy in
a plane-wave approach and special care must be payed to it.

The exchange energy per unit cell for a periodic system is defined as
\begin{equation}
E_{x} = - \frac{e^2}{N} \sum_{^{\bk v}_{\bkp  v'}}
                         \int \frac{\phi_{\bk v}^*(\br)\phi_{\bkp v'}(\br)
                                    \phi_{\bkp v'}^*(\brp)\phi_{\bk v}(\brp) }
                                       {|\br-\brp|}  d\br d\brp,
\end{equation}
where an insulating and non magnetic system is assumed for simplicity.
Integrals and wave-function normalizations are defined over the whole
crystal volume, $V=N\Omega$ ($\Omega$ being the unit cell volume),
and the summations run over all occupied bands and all N $\bk$-points
defined in the BZ by Born-von Karman boundary conditions.

Introducing {\it codensities}, $\rho_{^{\bkp, v'}_{\bk,
v}}(\br) = \phi^*_{\bkp, v'}(\br) \phi_{\bk, v}(\br)$, with Fourier
transform, $\rho_{^{\bkp, v'}_{\bk, v}}(\bk-\bkp+\bG)$, the above
expression can be recast in reciprocal space as
\begin{equation}
E_x = - \frac{ 4\pi e^2}{\Omega} \times
           \frac{\Omega}{(2\pi)^3} \int d\bq \sum_{\bG}
           \frac{ A(\bq+\bG) }{|\bq+\bG|^2}
\label{HF1}
\end{equation}
where an auxiliary regular function 
$ A(\bq+\bG) = \frac{1}{N} \sum_{\bk} \sum_{v,v'} 
| \rho{^{\bk-\bq, v'}_{\bk, v}}(\bq+\bG) |^2$ has been introduced.

A direct integration of Eq.~(\ref{HF1}) on the regular $\bq$-point
grid, defined by the $\bk-\bkp$ differences of the points used for the
wave-function BZ sampling, is problematic due to integrable divergence
that appears in the $\bq+\bG \rightarrow 0$ limit. The problem has been
addressed several time in the literature and many related schemes have
been proposed to treat it. Most eventually resort to a procedure, first
proposed by Gygi and Baldereschi,\cite{GygiBaldereschi} where an easily
integrable term, that displays the same divergence, is subtracted from and
separately added back to Eq.~(\ref{HF1}). Of the many proposed forms
\cite{GygiBaldereschi, Carrier, Kresse} we adopt here one of the simplest
that can be straightforwardly applied to any crystal structure: we simply
subtract and add a term $A(0) e^{-\alpha |\bq+\bG|^2}/ |\bq+\bG|^2 $
from the integrand. The integration of the resulting non divergent
term can then be performed by a summation over a regular grid of points
(either the full $\bq$-point grid defined from the $\bk$-point differences
or, in order to reduce computational cost, on  a subgrid of it), while
the divergent integration can be performed explicitly.

The final results is
\begin{widetext}
\begin{equation}
 E_{x} = - \left\{
           \frac{4\pi e^2}{N_\bq \Omega} \left[ {\sum_{\bq,\bG}}^{'}
           \frac{ A(\bq+\bG)} {|\bq+\bG|^2} + 
           \langle\langle
           \lim_{\bq\rightarrow 0} \frac{A(\bq)-A(0)}{\bq^2}
           \rangle\rangle \right]
           + D A(0) \right\}
\label{EHF2}
\end{equation}
with 
\begin{equation}
 D =  \frac{4\pi e^2}{N_\bq \Omega} \left[- {\sum_{\bq,\bG}}^{'}
             \frac{  e^{-\alpha |\bq+\bG|^2} } {|\bq+\bG|^2} +  \alpha \right]
             + \frac{4\pi e^2}{(2\pi)^3}  \sqrt{\frac{\pi}{\alpha}}.
\end{equation}
\end{widetext}
The symbol $\sum^{'}$ in the preceding formulas implies that the term
$\bq+\bG=0$ is excluded. Notice that with our simple choice the expression
for the divergence correction, $D$, is just (the reciprocal-space part of)
the Ewald sum for a single point-charge, periodically repeated according
to the super-periodicity defined by the $\bq$-point grid used. As long
as the parameter $\alpha$ is chosen so that the reciprocal space sum is
converged the result is independent of $\alpha$ and gives a correction
that decay very slowly with the inverse of the cell linear dimension, $L$.
This $1/L$ dependence of the singularity correction 
is for instance evident in Fig. 2 of Ref.~\onlinecite{Carrier}, although
the scaling law of the correction was not explicitly pointed out there.
In general, the error incurred on by neglecting this correction would
be so large that it is almost invariably included in the calculation.

From a closer analysis of Eq.~(\ref{EHF2}), however, it is
also clear that, even when the main effect of the divergence
has been taken into account by the Gygi-Baldereschi procedure,
another correction needs to be included to properly describe the
spherically-averaged limit, $\langle\langle\lim_{\bq\rightarrow 0}
\frac{A(\bq)-A(0)}{\bq^2}\rangle\rangle$, that cannot be calculated
directly for $\bq=0$.

It can be shown, see the Appendix, that this term is simply related to the
{\it gauge invariant spread} of the occupied manifold, as can be defined
in the theory of maximally localized Wannier functions. \cite{Wannier}
If this term is neglected (as to the best of our understanding it has
always been neglected so far in the literature) an error proportional
to this spread and inversely proportional to the cell volume and the
number of $\bq$-points included in the BZ summation is made. This might
be the reason for the reported slow convergence of the exchange energy
with respect to BZ integration, even for simple systems such as bulk 
Silicon\cite{Garcia} or Argon.\cite{Harl}

We have therefore included in the calculation of the exchange energy an
estimate of the limiting term, based on the assumption that the grid
of $\bq$-points used for BZ integration is dense enough that a coarser
grid, including only every second point in each direction would also
be equally accurate. Since the limiting term contribute to the
integral with different weight in the two grids one can estimate its value
as:
\begin{widetext}
\begin{equation}
\langle\langle\lim_{\bq\rightarrow 0}\frac{A(\bq)-A(0)}{\bq^2}\rangle\rangle  = 
\frac{1}{7}{\sum_{dense}}^{'} \frac{ A(\bq+\bG)} {|\bq+\bG|^2} -
\frac{8}{7} {\sum_{coarse}}^{'} \frac{ A(\bq+\bG)} {|\bq+\bG|^2}.
\end{equation}
\end{widetext}
We have verified that the convergence with respect to BZ integration is
generally improved. For instance for the exchange energy of solid Argon,
calculated from LDA wave-functions, a regular grid of 6x6x6 points was
sufficient for a convergence of the order of 0.1 mRy, while a much denser
12x12x12 grid was needed in Ref.~\onlinecite{Harl}. 
\subsection{Computational cost}
To discuss the computational cost of our implementation, let us denote
by $N_v$, and $N_c$ the number of valence and conduction bands, and by
$N_{pw\psi}$, and $N_{pw\chi_0}$ the number of plane waves used to represent
wave functions and Kohn-Sham response functions, respectively. The basic
operations in our implementation is the calculation of the linear density
response via DFPT technique.  For a norm-conserving pseudopotential, the
computational cost of a linear response calculation is essentially $N_v
\times (N_v \times N_{pw\psi}+N_{pw\chi_0}$). Since this calculation is
done repeatedly in the iterative diagonalization procedure, the total
computational cost must be multiplied by the number of eigenvalues
($N_{eig}$) that we want to calculate and the number of iterations
$N_{iter}$ in the iterative diagonalization. While the latter is
likely independent of the system  size, it is expected and in fact
there is evidence\cite{Wilson} that the former is proportional
to it. Therefore the total scaling of our approach is proportional to
$N_{pw\psi}N_v^2N_{eig}$ which grows as the fourth power of the system
size. In other implementations of ACFD formulas which are based on the
evaluation of the full response matrix, the most time-consuming operation
is the construction of the non-interacting Kohn-Sham response function
whose computational cost is proportional to $N_{pw\chi_0}^2N_vN_c$.
This means that the computational cost of these approaches also scales
with the fourth power of the system size.

Note however that $N_{pw\psi}$ is typically smaller than $N_{pw\chi_0}$
by an order of magnitude since $\chi_0$ relates to density responses and
perturbing potentials whose kinetic-energy cutoff is four times larger
than the one needed for wave-functions in the case of norm conserving
pseudopotentials and even more than that in the case of ultra-soft
pseudopotentials. Also the number of eigenpotentials that needs to be
computed is expected to be at least an order of magnitude smaller than the
size of the response function ($N_{pw\chi_0}$) and the number of iteration
$N_{iter}$ is unlikely to exceed $10$, especially if in the imaginary
frequency scan good starting trial eigenpotentials are taken from the
previous frequency in the list.  Our rough estimate shows therefore that
the number of operations involved in our method is from $100$ to $1000$
smaller than the one needed for other implementations. Although real
CPU-time obviously depends on many details in the realization of each
approach, we are confident that our approach allows us to implement ACFD
formulas in a very efficient way.

Moreover, our approach also has significant advantages due to its
iterative nature. Iterative solution of the generalized eigenvalue
problem (\ref{eigenchi}) will converge very rapidly if the initial
guess of the eigenpotentials are already close to the sought solutions.
As mentioned earlier calculated solutions for a given imaginary frequency
can serve as starting points for nearby frequencies. A similar behavior
can be expected also for calculation of response functions in nearby
$\mathbf{q}$--vectors although it may be more convenient in this case
to exploit the very easy parallelization of the $\mathbf{q}$--vector
summation involved in the correlation energy formula, Eq. (\ref{Ec2k}).
Finally, as already mentioned in the general formulation, the analysis
of the response function in terms of eigenmodes allows us to analytically
perform the coupling constant integration with a significant computational
saving.
\section{Application to selected systems}
\label{App}
\subsection{Bulk Si}
\label{SiBulk}
We have chosen bulk Si system as a testing ground for our implementation
of ACFD theory since the computational cost for the case of bulk Si is rather moderate, which is convenient for convergence checks, and there exist several published data that we can
compare with. As a first check, we have calculated the $20$ topmost
eigenvalues of the dielectric matrix at $\mathbf{q}= (0,0,0.01)
\frac{2\pi}{a}$ and compared our results with those reported in
Ref.~\onlinecite{Wilson} where the same quantities were calculated
by the explicit diagonalization of the RPA dielectric matrix and by
an iterative diagonalization procedure, named Projective Dielectric
EigenPotential method (PDEP), similar to ours.\cite{PDEP} 
The results reported in
Table \ref{Si-eps} show perfect agreement of the two methods among
each other and with the explicit calculation.

\begin{table}[t!]
\begin{ruledtabular}
\begin{tabular}{crrr}
~Index&~RPA\cite{Wilson} & PDEP\cite{Wilson} & present work\\
\hline
  1 & 14.7432 & 14.7538 & 14.7611 \\
  2 &  3.4231 &  3.4237 &  3.4238 \\
  3 &  3.3908 &  3.3914 &  3.3915\\
  4 &  3.3908 &  3.3914 &  3.3915\\
  5 &  3.3908 &  3.3914 &  3.3915\\
  6 &  3.3908 &  3.3914 &  3.3915\\
  7 &  3.3589 &  3.3596 &  3.3596\\
  8 &  2.4910 &  2.4925 &  2.4925\\
  9 &  2.4910 &  2.4925 &  2.4925\\
 10 &  2.4910 &  2.4925 &  2.4925\\
 11 &  2.4910 &  2.4925 &  2.4925\\
 12 &  2.4905 &  2.4920 &  2.4920\\
 13 &  2.4905 &  2.4920 &  2.4920\\
 14 &  2.4716 &  2.4721 &  2.4721\\
 15 &  2.1964 &  2.1972 &  2.1972\\
 16 &  2.1960 &  2.1968 &  2.1968\\
 17 &  2.1959 &  2.1966 &  2.1967\\
 18 &  2.1959 &  2.1966 &  2.1967\\
 19 &  2.1958 &  2.1966 &  2.1967\\
 20 &  2.1958 &  2.1966 &  2.1967\\
\end{tabular}
\end{ruledtabular}
\caption[\textsf{Top 20 eigenvaules of the dielectric matrix for an
eight-atom silicon cubic cell calculated via the present iterative method
and compared with the results reported in Ref.~\onlinecite{Wilson}.}]
{\label{Si-eps} 
Top 20 eigenvaules of the dielectric matrix for an eight-atom silicon
cubic cell calculated via the present iterative method and compared
with he results reported in Ref.~\onlinecite{Wilson} for RPA and for
projective dielectric eigenpotential method}
\end{table}

\begin{figure}[b]
\centerline{\includegraphics[width=0.45\textwidth]{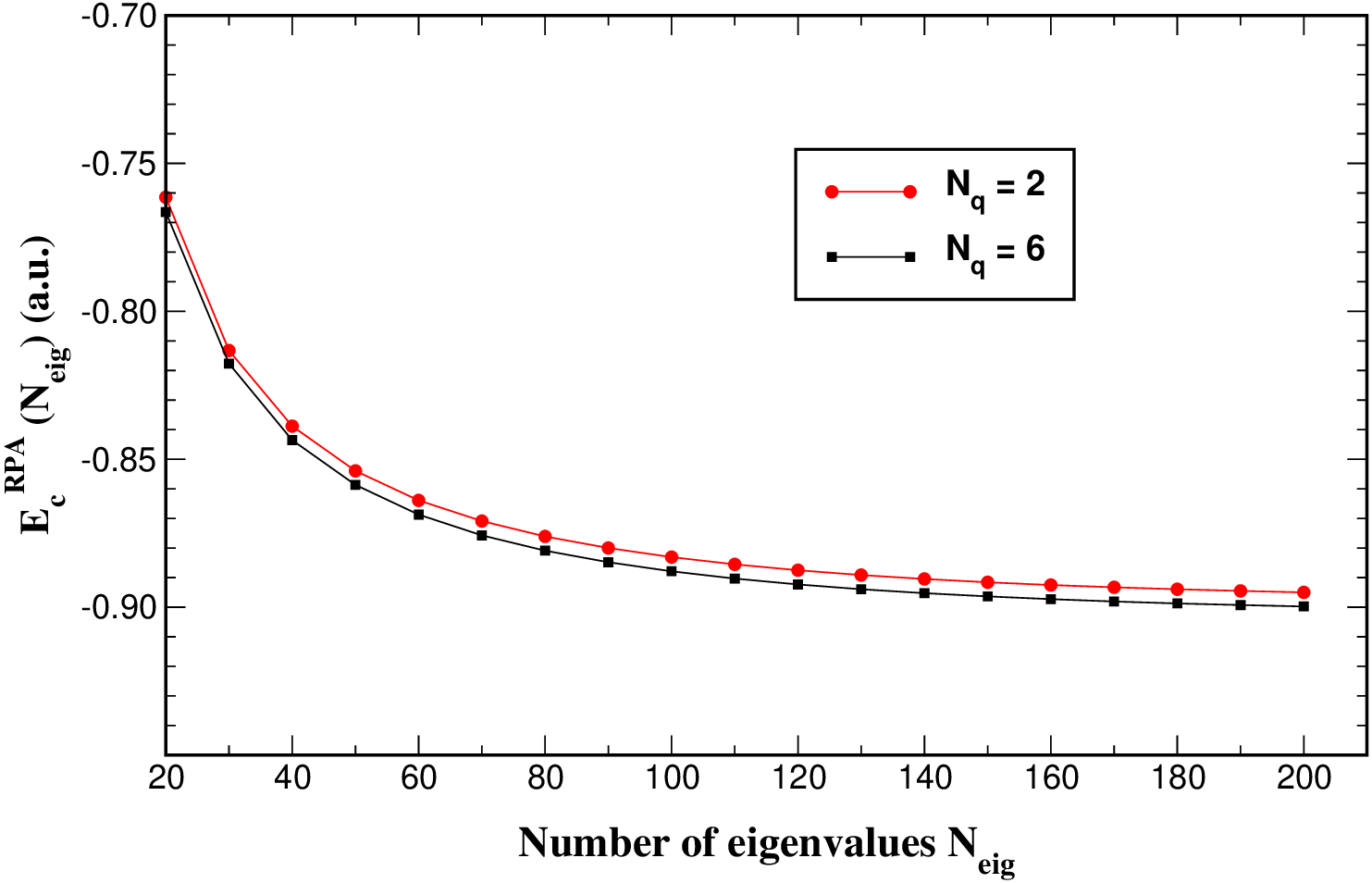}}
\caption{\label{fig:Ec_vs_nevl}
RPA correlation energy as function of the number of eigenvalues
included in the summation in Eq. (\ref{Ec2k}) for fcc bulk silicon. 
The two curves are for different number of special $\mathbf{q}$ points in 
the Brillouin zone integration of the same equation.}
\end{figure}

Let us now investigate the convergence of RPA correlation energy
with respect to the number of eigenmodes included in the summation of
Eq. (\ref{Ec2k}). 
To this purpose, we have used well-converged parameters
for standard LDA calculation of silicon ground-state charge density 
to evaluate the exact-exchange and RPA-correlation energies. 
The calculation was performed for the diamond structure in the
fundamental face-centered cubic cell with lattice constant of $10.20$
Bohr (corresponding to the theoretical equilibrium geometry at LDA level)
using a regular grid of $64$ k-point and a kinetic-energy cut-off of
$20$ Rydberg.

We show in Fig.~\ref{fig:Ec_vs_nevl} the dependence of RPA correlation
energy, $E_c^{RPA}$, on the number of eigenvalues, $N_{eig}$, included
in the summation. $E_c^{RPA}$ is indeed a rapidly converging function
of $N_{eig}$; truncating the sum after inclusion of $80$ or $100$
eigenvalues already ensures a convergence within a few tens of mRy.
Also the summation over special $\mathbf{q}$-points representing the
integration in the first Brillouin zone converges very rapidly; the
correlation energy changes only by a few mRy when the number of special
points increases from $2$ to $6$.

Next we compare our accurately calculated exchange and RPA(+) correlation
energies with the energies calculated within the pseudopotential
approximation but using quantum Monte Carlo techniques as reported
by Hood and co-workers in Ref.~\onlinecite{Hood}. Table \ref{SiQMC}
shows clearly that RPA alone gives indeed too negative values for the
correlation energy. When a simple local-density correction is added (RPA+),
the presently calculated correlation energy becomes much closer to the
Monte Carlo values. Obviously, the comparison would be more convincing
had we used the same pseudopotential used by Hood and co-workers.
Nonetheless, we believe that the good agreement between our calculations
and the results obtained by a very different method shows that our
results are at least in the right regime.

\begin{table}[t!]
\begin{ruledtabular}
\begin{tabular}{cccccc}
~~&~~&~$a_0$ ({\small bohr})~&~$E_x$ ({\small eV})~&~~&~$E_c$ ({\small eV})\\
\hline 
VMC~&~~&~- -~&~$-29.15$~&~~&~$-3.58 \pm 0.01$~\\
DMC~&~~&~- -~&~$-29.15$~&~~&~$-4.08 \pm 0.08$~\\\\
\multirow{2}{*}{EXX/RPA}
&~~&~ 10.20~&~ $-29.11$~&~~&~$-6.12$~ \\
&~~&~ 10.26~&~ $-28.98$~&~~&~$-6.11$~ \\\\
\multirow{2}{*}{ EXX/RPA+}
&~~&~ 10.20~&~ $-29.11$~&~~&~$-4.24$~ \\
&~~&~ 10.26~&~ $-28.98$~&~~&~$-4.23$~ \\
\end{tabular}
\end{ruledtabular}
\caption[\textsf{Exchange and correlation energies for bulk Si calculated by our method
compared to the values calculated by QMC techniques.}]
{\label{SiQMC} 
Exchange and correlation energies for bulk Si calculated by our method
compared to the values calculated using QMC techniques by Hood and co-workers
reported in Ref.~\onlinecite{Hood}.}
\end{table}

\begin{figure}[b]
\centerline{\includegraphics[width = 0.45\textwidth]{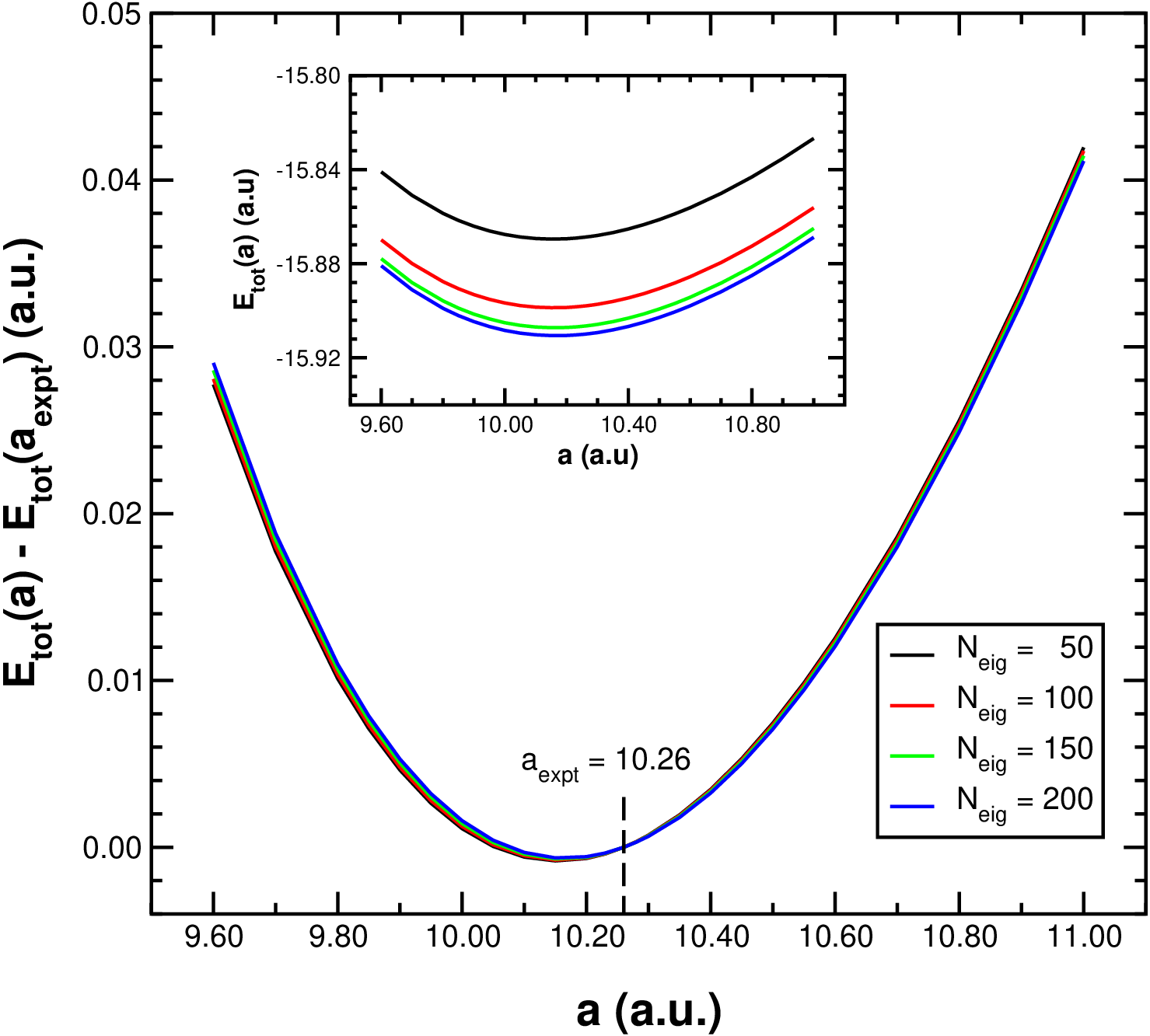}}
\caption{\label{Si_Etot_vs_lc_nevl}
{The total energy (per unit cell) differences at different lattice
constants and that at the experimental one of bulk silicon calculated
using different number of eigenvalues values included  in the summation
in Eq. (\ref{Ec2}) in our implementation of EXX/RPA+  scheme. The inset
is the true total energy. i.e. the values at the experimental lattice
constant is not subtracted.}}
\end{figure}

We have also calculated the total energy of bulk silicon as a function
of its lattice constant in EXX/RPA+ scheme in order to determine the
corresponding equilibrium properties.  Fig. \ref{Si_Etot_vs_lc_nevl}
shows the total energy per unit cell for bulk silicon as a function
of lattice constant calculated including different numbers $N_{eig}$
($N_{eig} = 50, 100, 150$ and $200$) of response function eigenvalues in
the summation in Eq. (\ref{Ec2}). For each curve the value calculated at
the experimental lattice constant is subtracted.  The resulting (nearly)
coincidence of the different curves confirms the expectation that energy
differences are rather insensitive to the number of eigenvalues
included in the RPA sum. This can also be appreciated in the inset, where
the unshifted EXX/RPA+ total energies are shown: as more eigenvalues
are included in the sum energy vs volume curves rigidly shift down and
although complete convergence is reached only when including about 200
eigenvalues already 100 or even 50 eigenvalues basically return the same
structural properties.

Table \ref{a0_Si} collect the predicted equilibrium lattice parameter
$a_0$, bulk modulus $B$ and pressure derivative of the bulk modulus $B'$
as a function of the number of eigenvalues $N_{eig}$ used to evaluate
RPA correlation energies. The changes in these quantities when increasing
$N_{eig}$ from $50$ to $200$ is very small, of the order of $0.1\%$ for
$a_0$ and $B$ and $1\%$ for $B'$. We can therefore conclude that only a
relatively small number of eigenvalues are needed in the evaluation of
ACFD RPA correlation energy to get a good description of many equilibrium
properties of the system.

\begin{table}[b] 
\begin{ruledtabular}
\begin{tabular}{ccccccc}
$N_{eig}$ &&  $a_0$ (a.u.)  &&  $B$ (GPa) &&  $B'$  \\
\hline
 50 && 10.155  &&  99.5 &&   4.22 \\
100 && 10.158  &&  99.4 &&   4.21 \\
150 && 10.162  &&  99.3 &&   4.19 \\ 
200 && 10.166  &&  99.1 &&   4.17 \\ 
~~&&~~ && ~~ && ~~\\
&LDA & 10.235  && 92.5 &&  4.16\\
&Expt\footnote{\scriptsize Experimental values are taken from Ref.~\onlinecite{Nielsen}} & 10.26  && 99.2 && 4.15 \\
\end{tabular}
\end{ruledtabular}
\caption{\label{a0_Si} 
Predicted equilibrium lattice parameter, $a_0$, bulk modulus, $B$,
and pressure derivative of the bulk modulus, $B'$, as function of the number
of eigenvalues $N_{eig}$ used to evaluate RPA+ correlation energies. 
The corresponding LDA and experimental values are also shown for comparison.}
\end{table}

Examining the predicted equilibrium lattice constants in Table \ref{a0_Si},
it seems that a more accurate (and sophisticated) treatment of xc-energy
in EXX/RPA+ scheme slightly worsens the agreement with experimental data
as compared with the LDA results. Nevertheless, as already pointed out
in Ref. \onlinecite{Garcia}, there are several points that can affect the
final results: (i) EXX/RPA+ scheme is applied in a non self-consistent
way using LDA Kohn-Sham orbitals, (ii) the pseudopotential itself has
been generated within LDA, and (iii) RPA+ is among the simplest possible
approximations to the xc-kernel within ACFD formalism. In view of these
shortcoming we believe that the present results for Silicon can be
considered satisfactory.

In spite of the improvement of numerical efficiency in our implementation,
we also observed a slow convergence of RPA correlation energy with respect 
to the kinetic-energy cutoff as reported in other
implementations.\cite{Marini,Harl}
Efficient extrapolation schemes that allow for the evaluation of correlation 
energies in the limit of infinite energy cutoff, already proposed for those 
implementations, can be easily adapted to ours. As for the convergence with
respect to BZ sampling, we notice, as explicitly shown in Ref.~\onlinecite{nhvietthesis}
for the case of bulk Si, that the difference between RPA correlation energies 
calculated using different number of $\bk$-points (for charge density) and 
$\bq$-points (for $E_c$) is a well-behaved function of the kinetic-energy 
cutoff, beyond a certain (not very large) value. This suggests an
extrapolation scheme, similar to the one proposed in Ref.~\onlinecite{Marini}, 
that might become useful for the calculation of more complex systems for which 
even the present very efficient implementation would be too demanding to reach 
complete convergence directly. The extrapolation procedure would be as follows: 
First a coarse grid of $\bk$- and $\bq$-point could be used for the 
calculation of RPA correlation energies at different kinetic-energy cutoffs.
Second, the corresponding correlation energy in the infinite cutoff limit 
could be obtained by extrapolating the results obtained at finite cutoffs. 
Finally, the errors due to coarse $\bk$- and $\bq$-point sampling of the BZ 
could be corrected by using finer grids evaluated at small kinetic-energy 
cutoff (whose safe value could be estimated from the convergence 
behavior of the correlation energy computed with the coarser grids.)
\subsection{Be$_2$ dimer}

Beryllium dimer is a paradigmatic example of the failure of
LDA/GGA DFT in the description of weakly bound systems. Previous
studies\cite{Richardson,Fuchs} have shown considerable discrepancies
between LDA/GGA and experimental results for binding energy, bond length
and vibrational frequencies. While the errors of LDA/GGA bond lengths
are in fact quite small (less than $2\%$), the vibrational frequency is
largely overestimated. The most severe discrepancy refers to the binding
energy: both LDA and GGA approximations overestimate the experimental
value by at least a factor of $4$. Fuchs and Gonze \cite{Fuchs} have
recently shown that EXX/RPA+ improves significantly the description
of Be$_2$ dimer over standard DFT. However, the error-bar of the RPA
correlation energy reported in that study is still as large as $30$~meV
which is of the same order of the experimental binding energy and the reported 
theoretical value. Our
efficient implementation of ACFD formulas allows us to reach a better
congergence, and thus to provide a better assessment of the performance
of EXX/RPA+ formalism for this system. To this end, we have carefully
checked the convergence with respect to all parameters involved in
our calculations, namely the kinetic-energy cutoff, the size of the
supercell used to simulate an isolated system using a periodic plane-wave
approach, and the number of eigenvalues included in the evaluation of
RPA correlation energy. 
\begin{figure}[t]
\centerline{\includegraphics[width=0.45\textwidth]{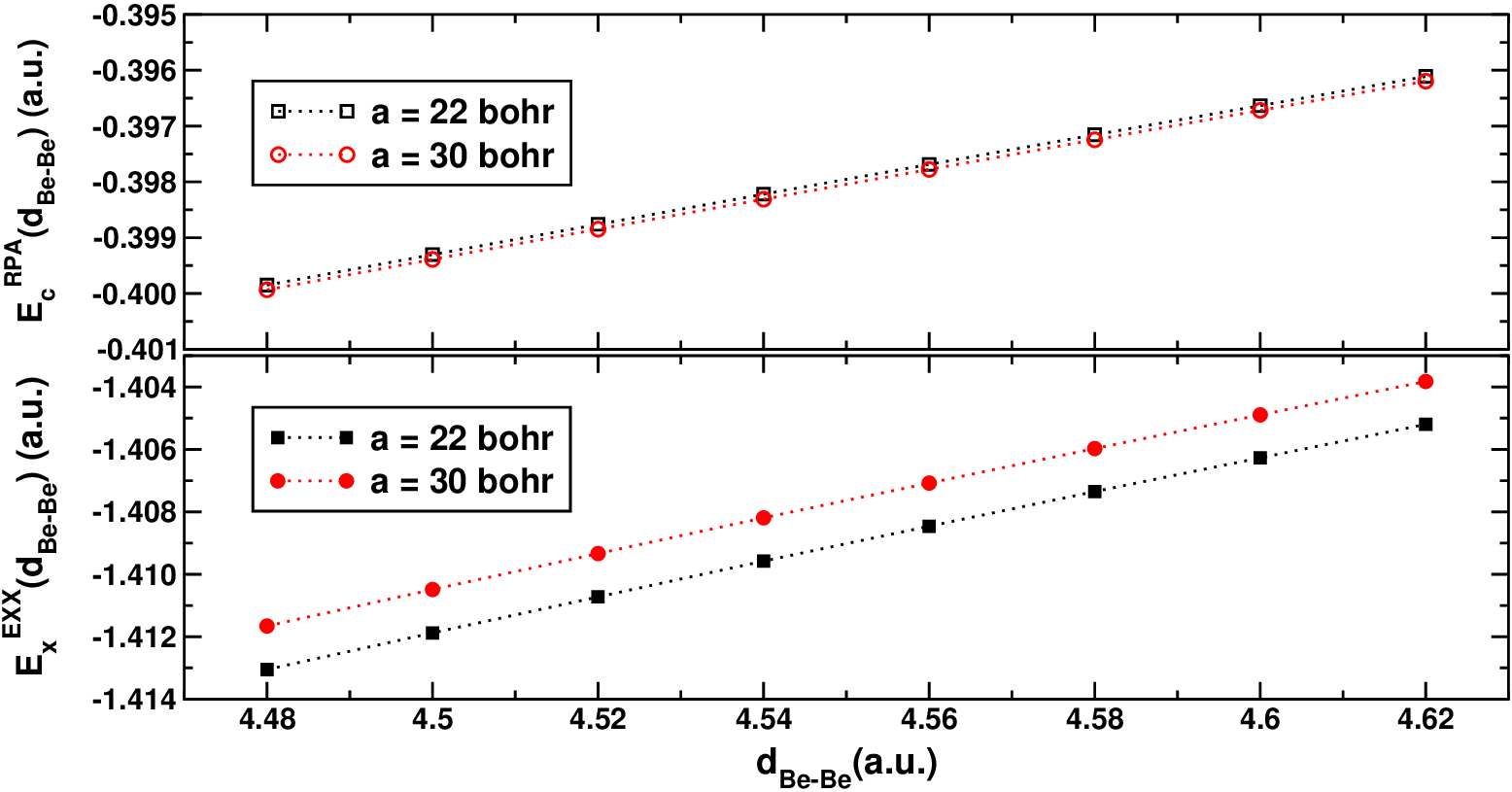}}
\caption{\label{ecex_vs_dd-lc}
Exact-exchange and RPA correlation energies of Be$_2$ as a function of
Be-Be distance ($d_{Be-Be}$) as calculated in a simple cubic cell of
side length of $22$ (square) and $30$ (circle) Bohr. Dotted lines are
simply drawn as a guide to the eye.}
\end{figure}

The same norm-conserving pseudopotential used
in Ref.~\onlinecite{Fuchs}, generated within exact-exchange Kohn-Sham
formalism, has been used here. It is a rather soft pseudopotential
and increasing the plane-wave kinetic-energy cutoff from $20$ to $30$
Ry makes the LDA total energy, the exact-exchange and RPA correlation
energies change less than $0.5$~mRy. This good convergence is consistent
with the use in Ref.~\onlinecite{Fuchs} of a plane-wave kinetic-energy
cutoff of $25$~Ry.

\begin{figure}[t]
\centerline{\includegraphics[width = 0.45\textwidth]{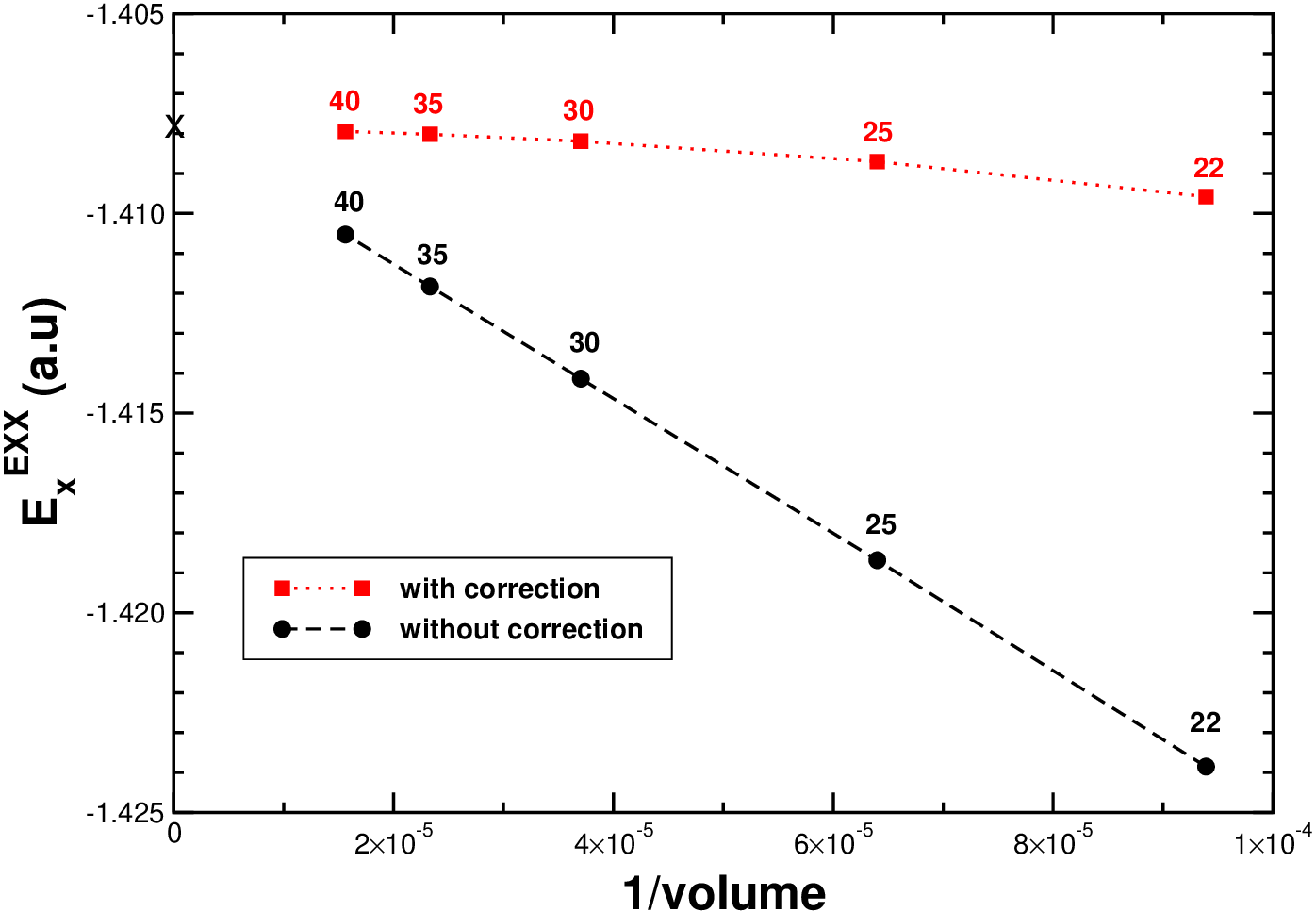}}
\caption{\label{ExBe2}
Exchange energy of Be$_2$ as a function of the supercell volume.
Both the calculation neglecting (filled circles) and including (filled
squares) an estimate of the $\mathbf{q}= 0$ term in the sum are reported.
Numbers close to the symbols correspond to the supercell linear dimension
in Bohr. } \end{figure}

Convergence with respect to the size of supercell is more
delicate. Figure~\ref{ecex_vs_dd-lc} shows exact-exchange and RPA
correlation energies as a function of Be-Be distances calculated
using supercells of $22$ and $30$ Bohr. While RPA correlation energy
varies by only a fraction of a mRy (top panel), the lower panel shows
a significant variation of several mRy for the exact-exchange energy.
As discussed in Sec.~\ref{exx_extra}, slow convergence of exact-exchange
energy with respect to supercell size--or the density of the BZ sampling
for an extended system--can result if, once the integrable divergence
is eliminated by the Gygi-Baldereschi procedure,\cite{GygiBaldereschi}
the residual $\mathbf{q}= 0$ term is not estimated correctly.  This is
shown in Figure~\ref{ExBe2} where the exact exchange energy of Be$_2$
molecule is shown as a function of the inverse supercell volume.  A large
error, proportional to the inverse supercell volume, is present when the
residual $\mathbf{q}= 0$ term is simply neglected, while a much better
convergence with system size is obtained when it is estimated according
to the recipe described in Sec.~\ref{exx_extra}.  While with a simple
cubic supercell of $22$ Bohr side the calculated exchange energy still
has an error of a couple of mRy, we are fully confident that using
a supercell of $30$ Bohr ensures a convergence of the exact-exchange
energy within  a fraction of mRy. One might expect 
some degree of error cancellation in energy differences when the residual
$\mathbf{q}=0$ terms are neglected in the exact-exchange calculations 
of Be dimer and of the separated Be atoms. 
Our explicit verification has shown 
that a variation of the order of mRy is still present in this difference 
when the supercell
size is increased from $22$ to $30$ bohr.

\begin{figure}[t]
\centerline{\includegraphics[width = 0.45\textwidth]{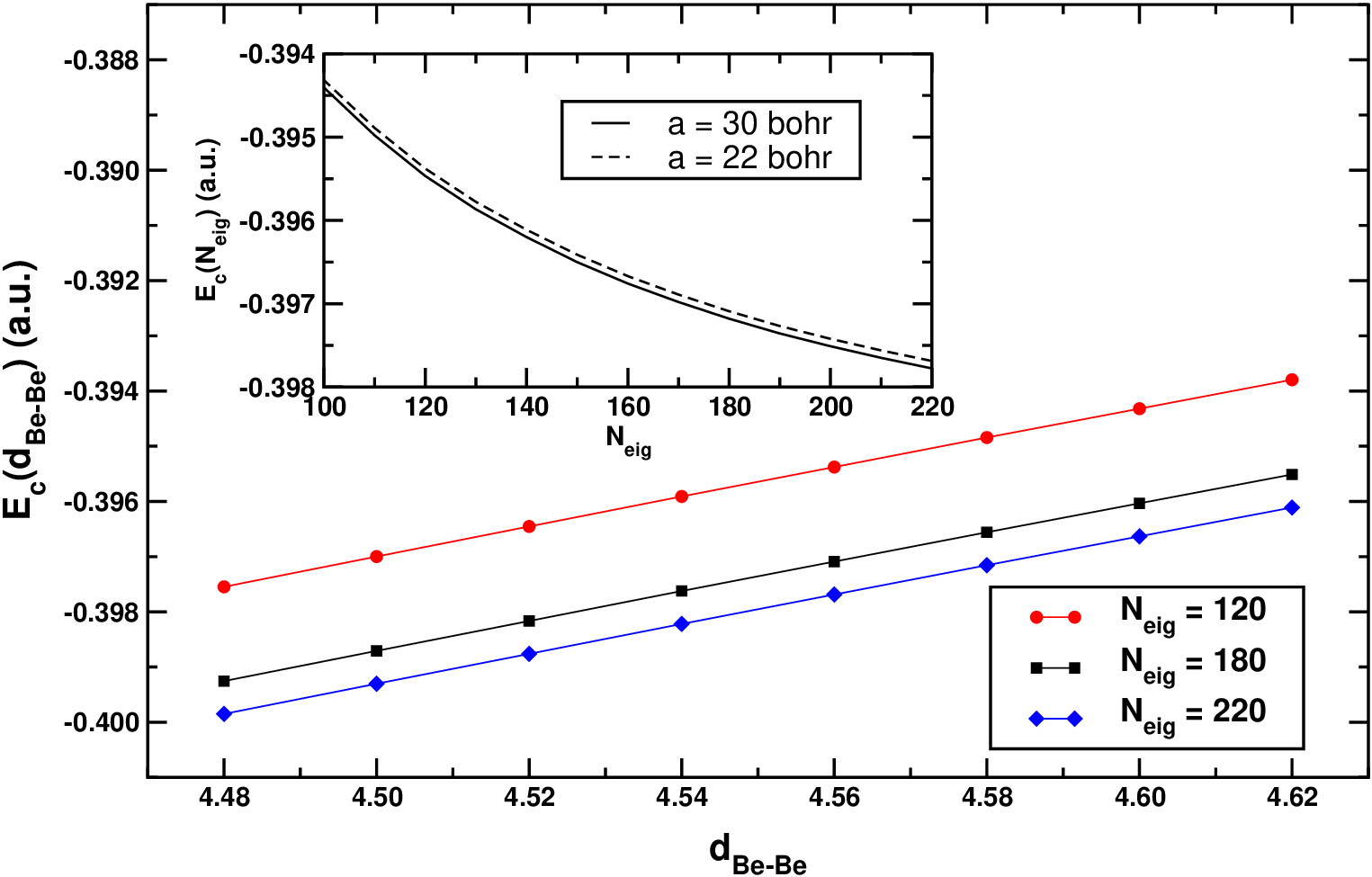}}
\caption{\label{ec_vs_dd-nevl}
RPA correlation energies of Be$_2$ at different Be-Be distances
($d_{Be-Be}$).  The curves are for different number of eigenvalues
$N_{eig}$ included for evaluation of Eq. (\ref{Ec2}): $N=120$ (circle),
$N=180$ (square), and $N=220$ (diamond).  The lines are simply drawn as
a guide. Inset: RPA correlation energies of Be$_2$ at the Be-Be distance
($d_{Be-Be}$) of $4.56$ Bohr placed in a simple cubic supercell with
the size length of $22$ (dashed line) and $30$ Bohr (solid line).}
\end{figure}

Finally, convergence of RPA correlation energy with respect to the
number of eigenvalues included in the ACFD summation in Eq.~(\ref{Ec2})
is shown in Fig.~\ref{ec_vs_dd-nevl} where RPA correlation energies of
Be$_2$ calculated using $120$, $180$, and $220$ eigenvalues are plotted
as a function of Be-Be distances. By including up to $220$ eigenvalues,
a convergence within $0.5$ mRy is obtained for the absolute value of
correlation energy. On the basis of the nearly parallel behavior of
the curves reported in Fig.~~\ref{ec_vs_dd-nevl} for different number
of included eigenvalues we can anticipate a much faster convergence for
energy differences, which actually determine
the equilibrium properties of the dimer.

\begin{table}[b]
\begin{ruledtabular}
\begin{tabular}{cllllll}
\multirow{2}{*}{$N_{eig}$}
& \multicolumn{2}{c}{E$_b$(eV)}
&\multicolumn{2}{c}{d$_0$(bohr)}&\multicolumn{2}{c}{$\omega_e$(cm$^{-1}$)}\\
\cline{2-3}\cline{4-5}\cline{6-7}
&{\tiny ~~~RPA}&{\tiny ~~~RPA+}&{\tiny ~RPA}&
{\tiny ~RPA+}&{\tiny ~RPA}&{\tiny ~RPA+}\\
\hline
 60 & -0.0667 & -0.0377 & 4.516 & 4.553 & 296.1 & 298.5 \\
120 & -0.0657 & -0.0368 & 4.521 & 4.558 & 296.6 & 298.7 \\
180 & -0.0655 & -0.0367 & 4.523 & 4.560 & 296.3 & 299.2 \\
220 & -0.0654 & -0.0365 & 4.524 & 4.561 & 297.1 & 298.5 \\\\
Ref. & -0.08(3) & -0.06(3) & 4.55~ & 4.59~ & 311 & 298 \\
Expt. & \multicolumn{2}{c}{-0.098}&\multicolumn{2}{c}{ 4.63}&\multicolumn{2}{c}{ 275.8}\\
\end{tabular}
\end{ruledtabular}
\caption{\label{a0Be2}
Equilibrium properties of Be$_2$ in EXX/RPA(+) scheme with different
numbers of eigenvalues included in the calculation of RPA correlation
energy.} 
\end{table}

Table~\ref{a0Be2} collects predicted binding energy, bond length,
and vibrational frequency of Be$_2$ calculated including different
numbers of eigenvalues in the ACFD evaluation of RPA correlation energy.
As expected, all these quantities are only very slightly changed when the
number of eigenvalues is increased from $60$ to a value as large as
$220$. Even the molecular binding energy appears to have a convergence
of the order of the meV when only 60 eigenvalues are included in the ACDF
sum. The result that properties related to energy differences are rather
insensitive to the number of eigenpotentials used in evaluating
RPA correlation energy, not only in the ``standard" case of silicon
(section~\ref{SiBulk}), but also in the much more delicate case of Be$_2$
dimer is very promising for the future application of the present method
to complex realistic systems.

When compared with the reference values reported in Ref.~\onlinecite{Fuchs},
our binding energies, both in RPA and RPA+, are systematically smaller,
although still in agreement within the large error bar of the previous
calculation.
Moreover, had we calculated the binding energy of Be$_2$ without the
$\mathbf{q}= 0$ correction described above, as done in the earlier
calculation,\cite{Fuchs-private} a much better matching with the ACFD
calculation in Ref.~\onlinecite{Fuchs} would have been obtained.

Coming to the comparison with the experimental results it is clear
that while EXX/RPA+ scheme definitely improves the poor performance of
LDA and GGA for weakly bound systems like Be$_2$, the results may not
be as good as suggested in Ref.~\onlinecite{Fuchs}. The performance of RPA+
and other ACFD-based schemes to describe realistic weakly bound systems
needs to be more systematically investigated.

\subsection{Spherical atomic systems}
\begin{figure}[b]
\centerline{\includegraphics[width=0.45\textwidth]{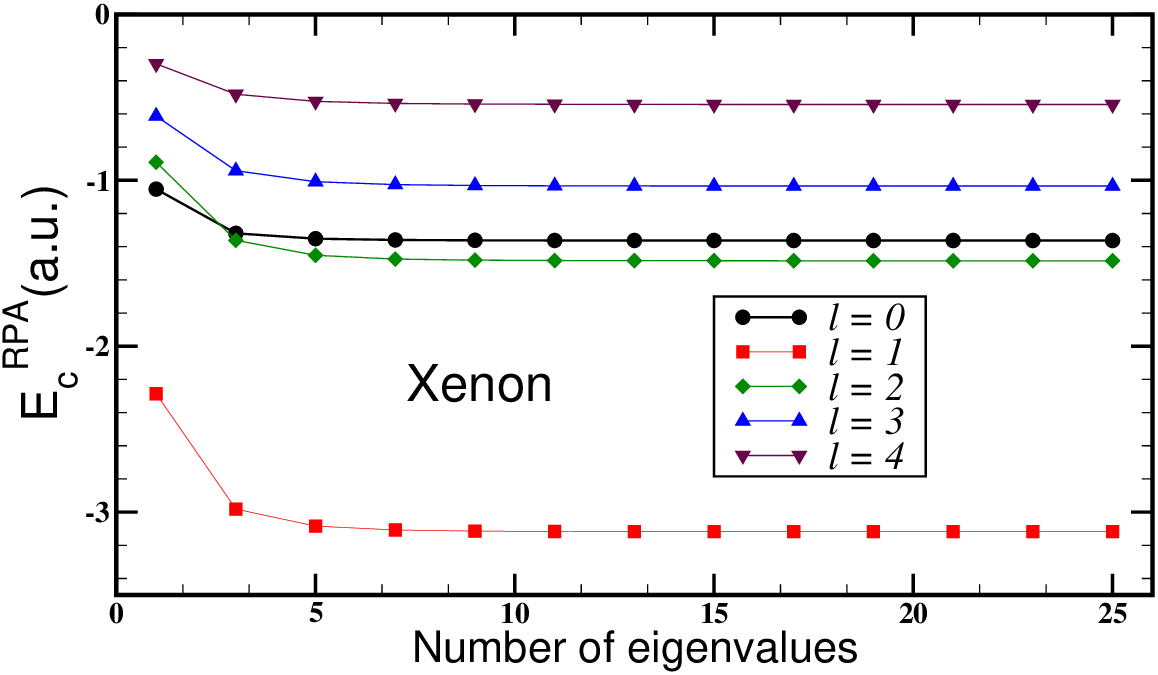}}
\caption{\label{fig:Ec-vs-Na}
The dependence of RPA correlation energy on the number of eigenvalues
for Xenon atom. The curves show the cumulative contribution to RPA
correlation energy for some of the lowest angular momentum numbers
(higher $l$ are not plotted). Including the first $15$ eigenvalues is
enough to ensure a convergence within $1mRy$, which is also used as the
threshold for convergence with respect to angular number $l$.}
\end{figure} 

\begin{table*}[t]
\caption{Full RPA and RPA+ correlation energy (in Rydberg atomic
units) of spherical atoms compared to the reference and exact values.
The reference data were calculated from EXX-only (i.e. exact-exchange and
no correlation) KS orbitals in a different implementation.\cite{Jiang}}
\begin{ruledtabular}
\begin{tabular}{clrcccccc}

%
Atom~&$~~E_c^{\textrm{\tiny expt}}$~&$E_c^{\textrm{\tiny LDA~}}$
&\multicolumn{3}{c}{$E_c^{\textrm{\tiny RPA}}$}&\multicolumn{3}{c}{$E_c^{\textrm{\tiny RPA+}}$}\\
\cline{4-6}\cline{7-9}
~~~~&~~~~&~~~~&{$\rho^{\textrm{\tiny LDA}}$}&{$\rho^{\textrm{\tiny EXX}}$}&~Ref.~\onlinecite{Jiang}&
{$\rho^{\textrm{\tiny LDA}}$}&{$\rho^{\textrm{\tiny EXX}}$}&~Ref.~\onlinecite{Jiang}\\
\hline
He~~&{-0.084}\footnote{\scriptsize~Experimental values quoted in Ref. \onlinecite{Kurth}. } &-0.229~~~&-0.168~&-0.167~&-0.166~&{-0.096}~~&{-0.094}~~&0.094\\
Be~~&{-0.190}\footnotemark[1]&-0.447~~~&-0.373~&-0.367~&-0.358~&{-0.230}~~&{-0.224}~~&0.216\\
Ne~~&{-0.786}\footnotemark[1]&-1.474~~~&-1.216~&-1.195~&-1.194~&{-0.821}~~&{-0.800}~~&0.800\\
Ar~~&{-1.463}\footnotemark[1]&-2.842~~~&-2.221~&-2.206~&-2.202~&{-1.503}~~&{-1.487}~~&1.482\\
Kr~~&{-4.15}\footnote{\scriptsize ~ Difference of ``exact" and HF total energies reported in Ref. \onlinecite{Ma}.} ~~&-6.533~~~&-5.226~&-5.192~&~n/a~&{-3.736}~~&{-3.702}~~&n/a\\
Xe~~&{-6.86}\footnotemark[2]~~&-10.358~~~&-8.312~&-8.278~&~n/a~&{-6.049}~~&{-6.016}~~&n/a\\
\end{tabular}
\end{ruledtabular}
\label{table:Full-RPA}
\end{table*}

Benchmark results of RPA correlation energies for a number of spherical
atoms calculated by constructing the full response function from
the spectrum of Kohn-Sham Hamiltonian have been recently reported in
Ref.~\onlinecite{Jiang}.

For spherically symmetric systems, ground state Kohn-Sham orbitals are
classified by their principal quantum number, $n$, and by angular momentum
numbers $l,m$. The KS equations can be solved numerically on a radial
grid, within a given approximate, LDA or GGA, functional.  Similarly, the
non interacting and interacting response functions are block-diagonal with
respect to angular momentum $l$, and $(2l+1)$-fold degenerate with respect
to $m$. Thus, contributions to the ACFD formula from different angular
momenta can be calculated independently and added up.  For each angular
momentum the calculation proceeds as follows. 
(i) A trial potential is selected and the corresponding linear
density response is calculated by solving the modified Sternheimer
equation.\cite{Mahan80} A single iteration, with no self-consistency,
is needed since the non-interacting response-function is studied.
(ii) The calculated density response is orthogonalized, with overlap
matrix $v_c$, with respect to any previously computed one. The generating
potential is accordingly transformed, exploiting the linearity of $\chi_0$.
By calculating the Hartree potential from the resulting density a new
trial perturbation is generated.
(iii) A matrix representation of $\chi_0$ on the hence generated
trial-potential basis is built, and diagonalized to get the
eigenvalues $a_i$'s.
This three-step process is repeated until convergence in the sum over
eigenvalues in Eq. (\ref{Ec2}) is reached. The same calculation procedure
is then repeated at different values of angular momentum, $l$, and imaginary
frequency, $iu$.\cite{GLscheme}

Let us investigate the convergence of the RPA $E_c$ with respect to
the number of eigenvalues of the generalized eigenvalue problem of
Eq.~(\ref{eigenchi0}). Fig. \ref{fig:Ec-vs-Na} shows the dependence
of RPA correlation energy--separated in the different angular momentum
contributions--on the number of eigenvalues included in the sum for Xenon
atom, the heaviest atom considered in this work (56 electrons). 
Basis set convergence is carefully checked and, for the case of Xe,
a basis-set size of $25$ trial potentials is enough for the desired
accuracy. It is clearly seen that the correlation energy converges
quite rapidly; including up to $15$
eigenvalues is enough in order to convergence the total correlation energy
within $1mRy$. These calculations therefore confirm explicitly, also
for the case of spherical atoms, our expectation that RPA correlation
energy can be obtained from only a small number of eigenmodes of the
non-interacting response-function of the system.

Table \ref{table:Full-RPA} shows the full RPA and RPA+ correlation
energies calculated with our method for a number of spherical atoms
whose ground state densities have been generated from EXX-only\cite{EXX}
and standard LDA functionals. Experimental\cite{Kurth} or accurate
theoretical\cite{Ma} values are also shown for comparison, together
with reference RPA and RPA+ correlation energies calculated recently,
\cite{Jiang} within a different implementation of the ACFD approach,
from EXX-only charge densities. All our calculated values in table
\ref{table:Full-RPA} are converged within a few mRy. The results slightly
depend on the quality of the ground-state density, i.e. on the different
approximate xc-functionals used in the self-consistent ground state
calculation. Focusing on the results obtained starting from EXX-only
charge densities, our calculated values for the full RPA correlation
energy (the fifth column) agree well with the reference data (sixth
column) within the error bar (with the exception of Be case); a similar
agreement is found for RPA+ correlation energies (eighth and ninth columns
of table \ref{table:Full-RPA}). The small residual differences between
the values obtained in the present and in the reference calculations
may probably be attributed to some slight residual difference in the
electronic densities used as input.

Not surprisingly, RPA correlation energies alone largely overestimate
the exact values. When combined with a local-density corrections to form
the RPA+ approximation the correlation energy compare very favorably
with the exact values thus giving support to the validity of RPA+ scheme.
\section{Conclusion}
In this work, we have proposed an efficient method for the calculation of
RPA correlation energies in the adiabatic-coupling fluctuation-dissipation
formalism.  Our approach involves the evaluation of a relatively
small number of eigenvalues of the non-interacting response function,
obtained combining concepts from density functional perturbation theory
with iterative diagonalization techniques.  General strategies as well
as technical details of the method as implemented in the plane-wave
pseudopotential Quantum-ESPRESSO distribution have been discussed to
some extent.

We have applied the method to study a few systems, representative of
bulk solids, weakly bound molecules and atoms. While the  study of
bulk silicon crystal helps validate the implementation, our study of
Beryllium dimer, thanks to its improved numerical accuracy with respect to
previous studies, allows us to gain a clearer picture of the performance
of EXX/RPA+ scheme in describing weakly bound systems.  Our calculation
confirms the important improvements of EXX/RPA+ with respect to LDA
or GGA but also shows that its performance in delicate cases can be
less impressive than previously concluded. The good agreement of RPA+
correlation energy with experimental or accurate theoretical results
for a few spherical atoms lend anyhow support to the quality of RPA+
correlation energies.

We are confident that the possibility of a careful control of the
numerical accuracy in ACFD calculation, resulting from the improved
numerical efficiency of our method, will turn out to be very useful in
the needed analysis of the performance of such a sophisticated density
functional in the study of many other realistic systems.

\acknowledgments
We thank M.\ Fuchs for providing us the pseudopotential used for calculations
in Ref.~\onlinecite{Fuchs} and for useful discussions.

\appendix
\section*{Appendix}
In this appendix we demonstrate that the spherically-averaged
limit, $\langle\langle\lim_{\bq\rightarrow 0}
\frac{A(\bq)-A(0)}{\bq^2}\rangle\rangle$, is simply related to the {\it
gauge invariant spread} of the occupied KS manifold, $\Omega_I$, defined
in the theory of maximally localized Wannier functions \cite{Wannier} as
\[
\Omega_I = \sum_{n} \left[ \langle \bz n| \br^2|\bz n\rangle -
\sum_{\bR,m} |\langle \bz n| \br|\bR m\rangle |^2 \right],
\]
where $\{ |\bR n\rangle \}$ is (any) set of Wannier functions
that describes the occupied manifold.

Starting from the explicit expression for $A(\bq)$ in terms of the
occupied Bloch states
\[
 A(\bq) = \frac{1}{N} \sum_{\bk} \sum_{v,v'} 
\langle\phi_{\bk,v}|e^{-i\bq\br}| \phi_{\bk-\bq,v'}\rangle
\langle\phi_{\bk-\bq,v'}|e^{+i\bq\br}| \phi_{\bk,v}\rangle,
\]
let us expand them in terms of the Wannier functions,
\[
|\phi_{\bk,v}\rangle =  \sum_n U_{v,n}(\bk) \sum_\bR e^{i\bk\bR} |\bR n\rangle,
\]
where $U_{v,n}(\bk)$ are general unitary transformations in the set of 
occupied bands at point $\bk$ in the BZ.
Inserting this definition in the expression for $A(\bq)$, this becomes,
after some straightforward manipulation,
\[
 A(\bq) = \sum_{\bR} \sum_{n,m} 
\langle\bR n|e^{-i\bq\br}| \bz m\rangle
\langle\bz m|e^{+i\bq\br}| \bR n\rangle.
\]
Expanding now the exponential factors in powers of $\bq$ and taking the
spherically averaged limit for $\bq\rightarrow 0$ one easily obtains
the desired result:
\[
\langle\langle\lim_{\bq\rightarrow 0}
\frac{A(\bq)-A(0)}{\bq^2}\rangle\rangle = -\frac{\Omega_I}{3} .
\]

\begin{thebibliography}{10}
\bibitem{GLscheme} 
For an efficient integration over imaginary frequencies
in Eq. (\ref{Ec2}), we divide the frequency axis into several
intervals and integrate within each interval by the Gauss-Legendre
method. Convergence within $1mRy$ is ensured by going to frequency high
enough that the absolute value of the integrand is as small as $10^{-8}$.
Such a tight threshold is very critical, especially for heavy atoms,
since frequencies as high as $10^6$ Ry must be considered and special
care must be taken in order to reach this accuracy. This is necessary
because the kernel in the correlation energy expression decays, roughly,
as the inverse of the sum of the squares of the imaginary frequency and
of the largest excitation energy in the system which is of the order of
thousands Rydberg for heavy atoms like Kr and Xe.  As for the summation
over angular momentum components, including up to $l=20$ is typically
enough to ensure the convergence within 1$mRy$.
\end{thebibliography}
\end{document}